# Dipole-moment induced capacitance in nanoscale molecular junctions

Ankur Malik, Ritu Gupta and Prakash Chandra Mondal *

Department of Chemistry, Indian Institute of Technology Kanpur, Uttar Pradesh 208 016, India

E-mail: pcmondal@iitk.ac.in (PCM).

**ABSTRACT**

Nanoscale molecular junctions are celebrated nanoelectronic devices for mimicking several electronic functions including rectifiers, sensors, wires, switches, transistors, and memory but capacitive behavior is nearly unexplored. Capacitors are crucial energy storage devices that store energy in the form of electrical charges. A capacitor utilizes two electrical conductors separated by a dielectric material. However, many oxides-based dielectrics are well-studied for integrating capacitors, however, capacitors comprised of thin-film molecular layers are not well-studied. The present work describes electrochemically grafted thin films of benzimidazole (BENZ) grown on patterned ITO electrodes on which a 50 nm Al is deposited to fabricate large-scale (500 x 500 $\mu m^2$) molecular junctions. The nitrogen and sulfur-containing molecular junctions, ITO/BENZ/Al act as a parallel-plate capacitor with a maximum capacitance of ~59.6 ± 4.79 $\mu Fcm^{-2}$. The present system can be an excellent platform for molecular charge storage for future energy applications.

# INTRODUCTION

Complementary metal oxide semiconductors (CMOS) play crucial roles to meet the demand for compact and fast electronics. However, it is believed by many that the microelectronic is about to reach the limits of miniaturization, which is constrained by both physical principles and production costs.[1] Molecular electronic is an important platform and can be an attractive and alternative to the CMOS. The molecular electronic deals with either a single or group of molecules sandwiched between two electrical conductors for understanding current-voltage (*I-V*) responses have been considered a potential building block for future nanoelectronics systems.[2–6] Charge transport through a single molecule, self-assembled monolayers (SAMs) or molecular thin films that are grown via electrochemical method is the basis of molecular electronics, which can mimic the electronic functions of traditional semiconductor devices such as diodes,[7] transistors,[8] sensors,[9,10] switches,[11,12] and memory.[13] Considering the modern days demand for high energy density, and high power density energy storage devices, inorganic oxides have been extensively explored.[14–17] However, such materials require high processing temperature, longer reaction time, and high-vacuum deposition. Besides, bottom contact is modified with either drop-casted or spin-coated which creates weaker electrode-materials interfaces and thus may suffer from long term stability. Conductive organic polymers such PANI and PANI/TCNQ composite thin films as capacitors are reported [18]. Molecules offer several advantages over oxide materials due to their low-cost and easy synthesis, tunable electronic properties, thermally stable, solution processability, forming covalent interfaces. However, towards this goal, molecular electronics as "on-chip" capacitors have not been explored so far, which could be the lack of proper molecular system design, and integration. Since molecules are nanoscale materials, a bottom-up approach in the fabrication of molecular electronic devices provides the advantage of controlling the thickness of molecules or films within the junctions. McCreery group reports charge storage devices using tetraphenylporphyrin, fluorene-benzoic acid heterostructures, and naphthalenediimide molecular layers placed between two carbon electrodes and the effect of solvent vapour and mobile ions on capacitance.[19–21] The same group also develops nanoscale molecular films that could increase capacitance by 100 folds (than unmodified carbon electrode) in acidic electrolyte.[22,23] However, all these demonstrations are limited to carbon electrodes fabricated via either electron-beam or PPF films which may not be available to all the researchers for further development. Besides, heteroatoms-containing molecules are known to exhibit high capacitance which have not been studies in the form of molecular junctions as 'on-chip' molecular capacitors. We attempt to address those issues by fabricating heteroatom containing molecular junctions that replaces carbon electrode with ITO. In doing so, we adopt the electrochemical diazonium reduction (E-Chem grafting) method which is seemingly popular among the other thin-films preparation.[24,25] We consider 5-amino-2-mercaptobenzimidazole for which several heteroatoms such as nitrogen, sulfur can facilitates increase electrical polarization which remains unexplored in solid-state molecular junctions for mimicking on-chip capacitance.

**RESULTS AND DISCUSSION**

The amino group was diazotized (BENZ-D) using NaNO$_2$ followed by structural confirmation via $^1$H-NMR (**Figure S1, ESI†**), FT-IR, and UV-Vis spectroscopy. The diazonium salt had an intense yellow colour while that of -amino-2-mercaptobenzimidazole was almost colourless (**Figure 1a**). UV-vis spectra of 5-amino-2-mercaptobenzimidazole and diazonium salt in DMSO were recorded on JASCO UV-Visible-NIR V-770 spectrophotometer. Two intense peak at 335 and 266 nm were observed in UV-vis spectrum of 5-amino-2-mercaptobenzimidazole. A red shift was observed as a broad band in range of 535-350 nm was present in UV-vis spectrum of diazonium salt along with peak at 286 nm (**Figure 1b**). The FT-IR spectrum of 5-amino-2-mercaptobenzimidazole and diazonium salt was collected using a Bruker Alpha-II instrument set to ATR mode and measuring wavelengths between 500 and 4000 cm$^{-1}$. BENZ-D had an intense peak at 2250 cm$^{-1}$ which were absent in amino precursor confirming the formation of aryl diazonium group (Ar-N$^{2+}$). The peaks at 2560, 1620, 1460, 1020 cm$^{-1}$ are attributed to S─H stretching, C═C, C─H bending and B─F stretching (**Figure 1c**). A 100 nm thick and 500 µm width ITO electrodes were fabricated on quartz substrate via PDC Magnetron sputtering using a custom-made shadow mask (**Figure S2**). Freshly prepared diazonium salt solution (4 mM) was electrochemically reduced on chemically modified OH-terminated ITO (as a working electrode in three electrodes set-up, **Figure 2a**) to produce strong, high yield, fast forming, robust covalent bonding between the molecule and the patterned ITO substrates. It is possible to grow molecular layers of desired thickness thanks to the E-Chem method. Electrochemical reduction of diazonium salt followed by thin films formation was performed by sweeping the working electrode potential from 0 V to -0.85 V (vs. Ag/AgNO$_3$) for up to 6, 9, 12, and 15 repetitive CV scans (Films 1-4, respectively) recorded at 0.1 Vs$^{-1}$ (**Figure 2b**) for 15 scans, others are in SI, **Figure S3a-c**). The voltammogram clearly manifests aryl diazonium reduction peak forming reactive radicals at -0.64 V and seems most of the diazonium salts accessible near the ITO surface are consumed for molecular films formation. Surface coverage of Films-4 was estimated at $6.32 \times 10^{-10}$ mol cm$^{-2}$ (The grafting conditions and surface coverage of Films 1-4 is given in **Table S1**) which is in good agreement of well molecular density on ITO surface. The morphology of Films-4 was analyzed using non-contact mode AFM showing an increased roughness of $3.7 \pm 0.2$ nm (**Figure 2c-d**). Such enhancement of roughness as compared to the bare ITO ($R_{rms}$ ~$2.8 \pm 0.2$ nm) is common for E-Chem grafted molecular layers, hence our finding confirms formation of nanometric BENZ layers on ITO. The thickness a Films 1-4 was calculated by AFM in non-contact mode, Films-4 had the highest thickness of $23.0 \pm 1.7$ nm (**Figure S4-S5, Table S1**).

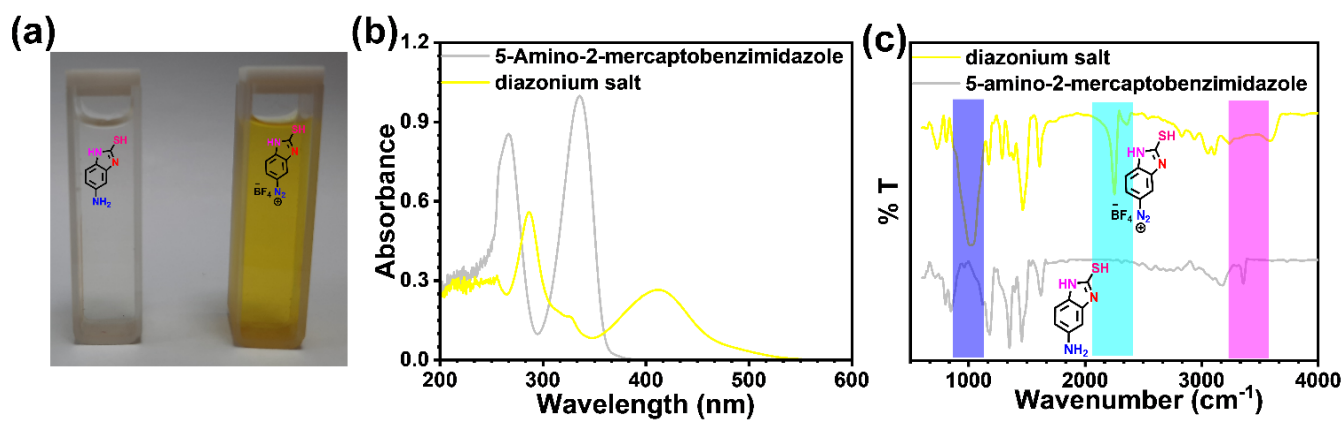

**Figure 1.** (a) Digital photograph of solutions of 5-amino-2-mercaptobenzimidazole (left) and diazonium salt (right), (b) comparison of UV-vis spectra and (c) comparison of FT-IR spectra of 5-amino-2-mercaptobenzimidazole (gray) and diazonium salt (yellow).

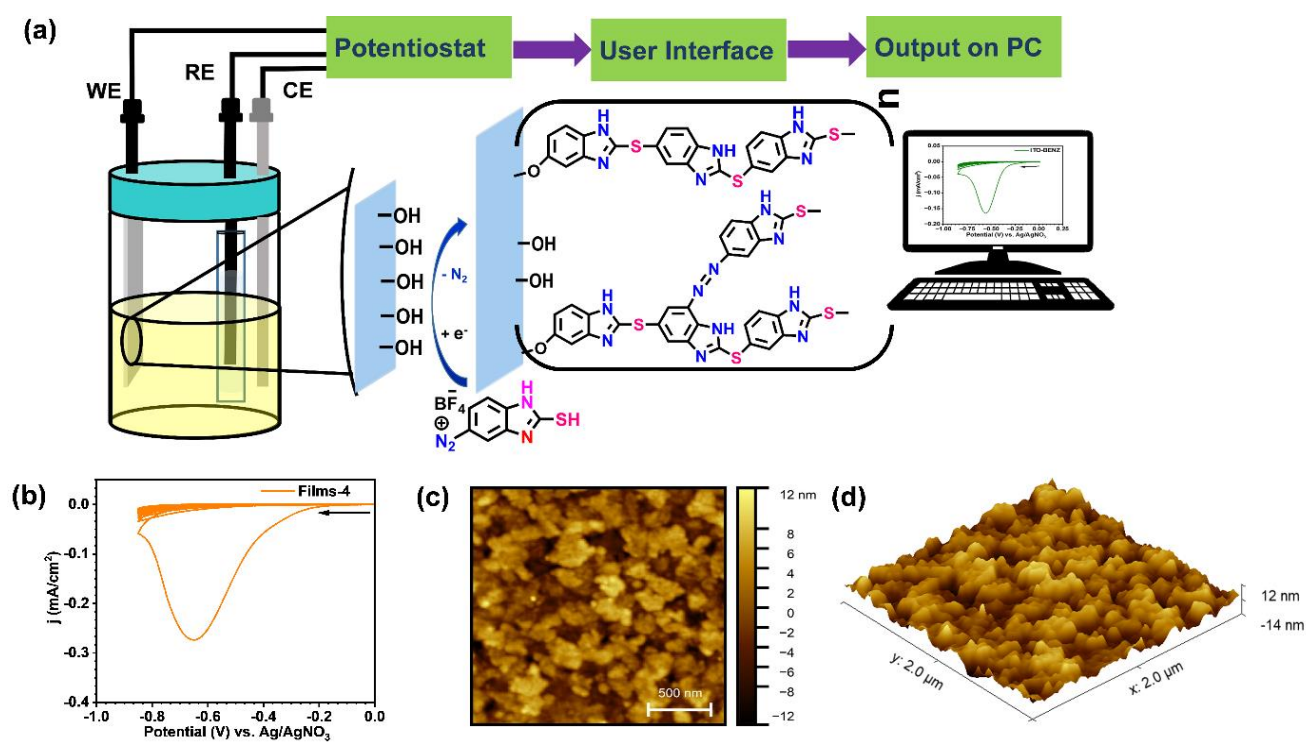

**Figure 2.** (a) The three-electrode electrochemical system for the reduction of BENZ-D, (b) cyclic voltammogram of E-Chem grafting of BENZ molecular layers on ITO electrodes (Films-4) and (c-d) 2 D and 3D AFM image of Films-4.

Thin film X-ray diffraction (XRD) pattern recorded on BENZ modified ITO electrode exhibited peaks at 2 theta values of 30.34, 35.33, 50.73, and 60.38° corresponding to the (222), (400), (441), and (622) planes of cubic indium tin oxides (JCPDS 00-039-1058) whose intensity got reduced than that of bare ITO (**Figure S6**). Since the films are grown via reactive radicals formation which is difficult to grow the assemblies in a specific pathway, thus PXRD suggests amorphous nature of BENZ molecular layers. Absorption band in the range of 450-620 nm and at 335 nm ensures can be seen in the UV-vis spectra of BENZ molecular layers on ITO (**Figure S7a**). A raise in absorbance at 335 nm from Films-1 to Films-4 was observed suggesting more random or sidewise growth of molecular films with increase in E-Chem grafting cycles (**Figure S7b**). The expended and deconvoluted X-ray photoelectron spectra (XPS) of constituent elements for bare ITO is presented in **Figure S8**, and that of Films-4 is given in **Figure 3** and results are summarized in **Table S2**. The standard XPS peaks of bare ITO constituents' elements i.e. $In5d_{5/2}$, $In5d_{3/2}$, $Sn5d_{5/2}$, and $Sn5d_{3/2}$ were observed 444.18, 451.70, 486.29 and 494.77 eV, respectively. The O1s peak can be deconvoluted into three peaks at 529.64, 530.32 and 531.23 eV corresponding to oxygen in oxide lattice with and without oxygen vacancies, and O─H bond, respectively.[26] Similarly, C, N, O, S, In, and Sn elements were detected in XPS spectrum of ITO-BENZ. The C1s peak can be fitted into four peaks at 284.63, 285.47, 286.22 and 287.15 eV for C─C/C═C, C═N C─S and C─N, respectively (**Figure 3a**), which are well-correlated with the previous report.[27] Nitrogen and sulphur were absent in bate ITO sample but clear XPS peaks of N1s and S2p were observed in ITO-BENZ XPS spectrum, which confirmed the presence of nitrogen and sulphur. N1s peak can be further fitted into three peaks at 398.44, 399.54, and 400.41 eV corresponding to pyridine-type nitrogen, pyrrolic-N and –N=N, respectively (**Figure 3b**).[28] O1s can be deconvolated into three peaks at 530.26, 531.93 and 532.93 eV corresponding to oxygen in oxide lattice with without oxygen vacancies, O─H and C─O, which confirmed In/Sn─O─C bonding (**Figure 3c**).[26-29] S2p can be deconvoluted into two peaks one at 163.59 eV for $S2p_{3/2}$ and another at 164.62 eV for $S2p_{1/2}$ (**Figure 3d**). $In5d_{5/2}$, $In5d_{3/2}$, $Sn5d_{5/2}$, and $Sn5d_{3/2}$ were observed 444.64, 452.17, 486.74 and 4945.28 eV, respectively, with slightly higher binding energy than bare ITO (**Figure 3e-f**).[30] Also, the intensity of O1s, In3d and Sn3d in ITO-BENZ was significantly reduced, confirming the formation of BENZ molecular layers on ITO electrodes. A 50 nm aluminium (Al) top electrode was deposited in crossbar fashion guided by mask (**Figure 4a, Figure S9a-b**). Two probes *I-V* measurements were performed to characterise electrical response of the molecular junctions (**MJ-1**, **MJ-2**, **MJ-3** and **MJ-4**).

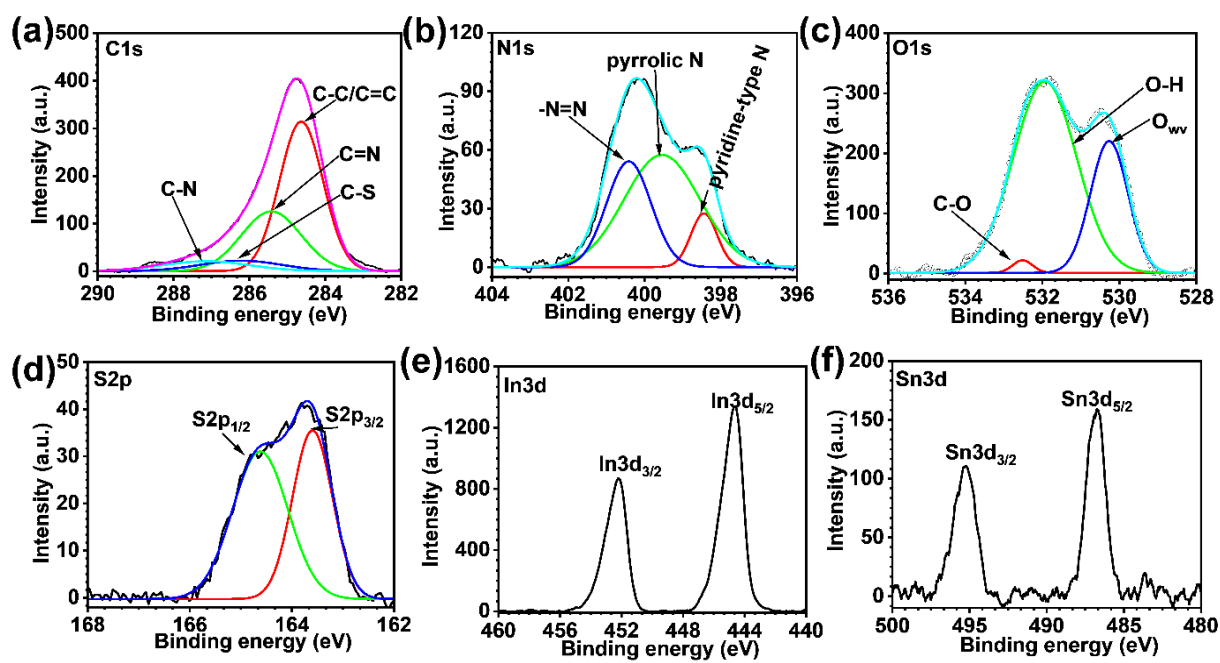

**Figure 3.** High-resolution and deconvoluted XPS spectra of (a) C1s, (b) N1s (c) O1s, (d) S2p, (e) In3d and (f) Sn3d of Films-4.

The *I-V* measurement was performed at various junctions from -2 to +2 Volts (the measurement setup **Figure 4b**) for all four fabricated devices (MJ-1 to MJ-4) and *j-V* plots (*j* = current density in mA/cm$^2$) are given in **Figure 4c**. The non-symmetrical sigmoidal *j-V* curves was observed in all devices, which might be due to two asymmetrical electrodes ITO and Al. The transport mechanism corresponds to multistep tunneling through a barrier defined by the difference in the LUMO and HOMO energies of the 2-mercaptobenzimidazole monomer.[21] A maximum current of -2.39, -0.84, -0.61 and -0.48 µA was obtained in negative bias, and 0.712, 0.28, 0.21 and 0.19 µA in positive bias for MJ-1, MJ-2, MJ-3 and MJ-4, respectively, which is consisted with the increase in thickness of molecular layers from MJ-1 to MJ-4. Another interesting feature was observed that at 0 V, current was not zero which can be clearly seen in *log j-V* plot in **Figure 4d** suggesting charge storing behaviour of BENZ molecular layers. The conductance vs voltage plots (**Figure 4e**) concluded that the device MJ-1 showed a sharp rise in conducting behaviour at -0.8V due to matching of energy level of BENZ molecular layers and one of the electrodes, while no such raise in conductance was observed in MJ-2, MJ-3 and MJ-4 suggesting these devices to be fully capacitive.

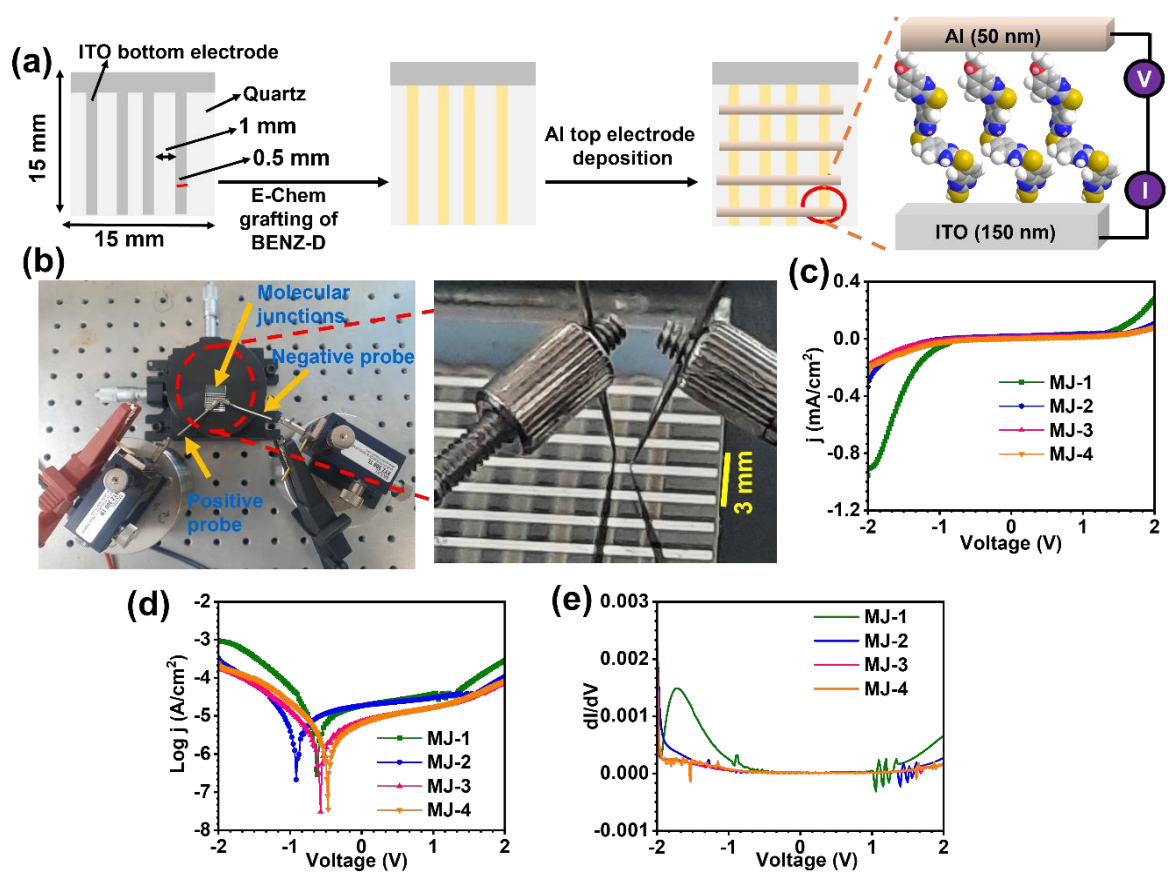

**Figure 4.** (a) Fabrication of ITO/BENZ/Al molecular electronic devices, (b) electrical measurements set-up, (c) *j-V* plot, (d) log *j-V* and (e) dI/dV vs V plots of MJ-1 to MJ-4.

Almost Ohmic behaviour is seen at the ITO/Al junction without forming any hysteresis (**Figure S10a-b**). The scan rate dependence (from 5 to 1000 mVs$^{-1}$) *I-V* loop behaviour of the ITO/BENZ/Al junctions were examined at room temperature in − 1 to + 1.5 V range. All four devices exhibited a *j-V* hysteresis

(**Figure S11**). With increase in the scan rate enclosed area also continued to increase suggesting a parallel-plate behaviour of molecular junctions and the charging and discharging halves of MJ-4 *j-V* hysteresis at 1Vs$^{-1}$ is given in (**Figure 5a**). The areal capacitance was calculated from area enclosed by *j-V* hysteresis at 10 mVs$^{-1}$. The calculated capacitances were 24.7 ± 0.98, 44.4 ± 3.1, 45.2 ± 2.26 and 59.6 ± 4.79 µFcm$^{-2}$ for MJ-1, MJ-2, MJ-3 and MJ-4 at a scan rate of 10 mVs$^{-1}$ (**Figure 5b**). The stability of junction was analyzed by running 1000 continuous *j-V* loop in MJ-4 and junction was stable and maintained and retained ~ 97% of initial capacitance (**Figure 5c, Figure S12**).

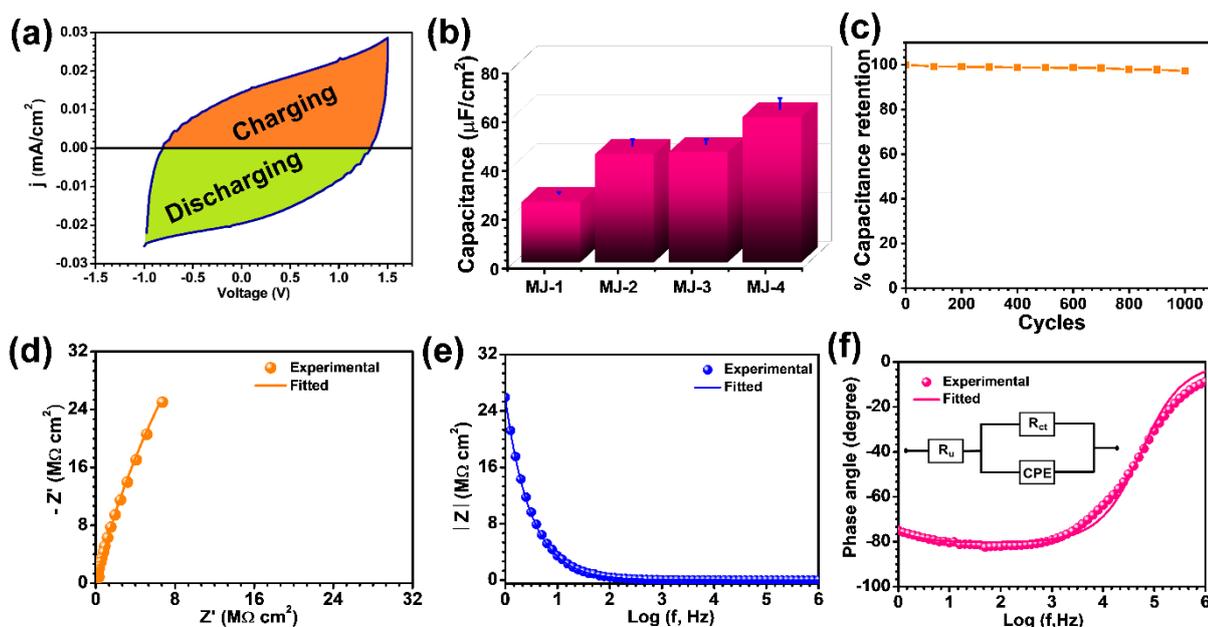

**Figure 5.** (a) Scan rate dependent *j-V* hysteresis in MJ-4 at 1 1 Vs$^{-1}$, (b) Capacitance of MJ-1, MJ-2, MJ-3 and Mj-4 at 10 mVs$^{-1}$, (c) capacitance retention in MJ-4 up to 1000 cycles, (d) Nyquist plot, (e)&(f) Bode plot of MJ-4 at V= 0 DC voltage in the frequency range 10$^6$ to 1 Hz with 100 mV amplitude.

Apart from DC studies, the capacitance was further evaluated by AC measurements. The MJ-4, at high frequencies, Nyquist plots showed a small semicircle illustrating a low charge transfer resistance ($R_{ct}$) of the junction (see The EIS measurement of ITO/Al junction **Figure S13-14**). Additionally, a straight line in a low-frequency regime parallel to the imaginary axis signifies the capacitive nature of the junction (**Figure 5c**). The bode plot also showed a phase angle of ~ -80° at 1 kHz, close to -90° for an ideal capacitor (**Figure 5e-f**)). An equivalent Randle's circuit model was used to determine all the circuit elements (**Figure 5e inset**)).[3]

In summary, we developed the BENZ molecular layers based capacitive molecular junction in ITO/BENZ/Al configuration. The thickness and hence capacitance of molecular junctions can be controlled by just varying the CV cycles during electrochemical grafting. The higher electron densities due to availability of lone pairs on heteroatoms in BENZ layers may hold more positive charge and can be possible reason for higher capacitance. On-chip charge storage in solid-state micro and nanoelectronics has always been challenging. The current study can inspire development of variety of heteroatom containing molecular electronic devices for nanoelectronics applications.

## ASSOCIATED CONTENT

**Supporting Information**

Reagent and Materials, Synthesis and characterization of diazonium salt of 5-amino-2-mercaptobenzimidazole, ITO bottom electrode patterning, E-Chem grafting procedure, AFM measurements, XRD patterns of bare ITO and Films-4, UV-vis spectra of Films 1-4, XPS measurements, Top electrode deposition, *I-V* measurement and capacitance calculation, Electrochemical Impedance Spectroscopy (EIS), References.

- **Conflicts of interest**

No conflicts of interest to declare for this work.

## ACKNOWLEDGMENTS

This work was financially supported by the Council of Scientific & Industrial Research, project NO.:01(3049)/21/EMR-II, New Delhi, India, and Science and Engineering Research Board (Grant No. CRG/2022/005325), and IIT Kanpur (IITK/CHM/2019044). AM, RG thank IIT Kanpur for an Institute post-doctoral fellowship (PDF254) and senior research fellowship, respectively. Authors thank IIT Kanpur for infrastructure and equipment facilities.

## DATA AVAILABILITY

All the data related to this work are available upon request to the corresponding author.